\documentclass[twocolumn,superscriptaddress,aps,preprintnumbers]{revtex4-1}

\usepackage{graphicx}
\usepackage{bm}
\usepackage{color}
\usepackage{epstopdf}
\usepackage{amsmath}
\usepackage{amssymb}
\usepackage{epstopdf}

\usepackage[urlcolor=blue,colorlinks=true,citecolor=blue,linkcolor=blue,pdfstartview={FitH},bookmarks=false]{hyperref}

\graphicspath{{fig/}{./fig/}{.}}

\begin{document}

\title{Leakage of the Majorana quasiparticles in Rashba nanowire deposited on superconducting--normal substrate}

\author{Aksel Kobia\l{}ka}
\email[e-mail: ]{akob@kft.umcs.lublin.pl}
\affiliation{Institute of Physics, M. Curie-Sk\l{}odowska University, \\ 
pl. Marii Sk\l{}odowskiej-Curie 1, PL-20031 Lublin, Poland}

\author{Andrzej Ptok}
\email[e-mail: ]{aptok@mmj.pl}
\affiliation{Institute of Nuclear Physics, Polish Academy of Sciences, \\ 
ul. E. Radzikowskiego 152, PL-31342 Krak\'{o}w, Poland}

\date{\today}

\begin{abstract}
Recent experiments show the possibility of realization of the Majorana quasiparticles at the end of the low dimensional structures. 
In this type of systems, interplay between spin orbit coupling, superconductivity and magnetic field leads to the emergence of the Majorana bound states in the topologically non-trivial phase.
Here, we study the nanowire located partially at the normal and superconducting base, using microscopic model of this structure and the Bogoliubov--de~Gennes technique.
We discuss the possibility of the \textit{leakage} of the Majorana bound state, located at the part above superconducting substrate to the part above normal material.
We have shown that this is possible only for some specific potential applied to the normal part.
\end{abstract}


\maketitle

\section{Introduction}
\label{sec.intro}

Majorana bound states (MBS) are an emergent phenomena existing in solid state physics. Since its experimental validation, there has been a shift of interest in the scientific community, as the proposed ideas regarding realization of Majorana quantum computers started to look promising~\cite{nayak.simon.08,aasen.hell.16,karzig.knapp.17,hoffman.schrade.2016}. 
Recently, realisation of the MBS have been reported in hybrid semiconductor--superconductor nanowire~\cite{mourik.zuo.12,deng.yu.12,das.ronen.12,deng.vaitiekenas.16,niechele.drachmann.17} and in ferromagnetic chain at the superconducor surface~\cite{nadjperge.drozdov.14,pawlak.kisiel.16,jeon.zie.17}.
Being a topological state, MBS robustness against external influence is essential in overcoming the problem of decoherence of quantum state, one of the main halting points in realization of quantum computing~\cite{albrecht.higginbotham.16}.
Implementation of such system requires three main ingredients: induced superconductivity, strong spin-orbit interaction and external magnetic field. 
Together, all of the above result in topological phase shift to non--trivial phase as the Cooper pairs pairing type changes from {\it s-wave} to {\it p-wave}~\cite{zhang.tewari.08,quay.hughes.10,sato.takahashi.10,seo.han.12,heedt.traverso.17}. 
This is due to the pairing of electrons from different Rashba bands, therefore having non-opposite momentum ${\bm k}$.
As a result, this allows for Andreev bound states (ABS) residing inside superconducting gap to coalesce into MBS on zero energy under conditions mentioned above~\cite{chevallier.sticlet.12,chevallier.simon.13,liu.sau.17b,ptok.kobialka.2017,prada.aguado.17,escribano.yeyati.17}.

\begin{figure}[!b]
\centering
\includegraphics[width=0.7\linewidth]{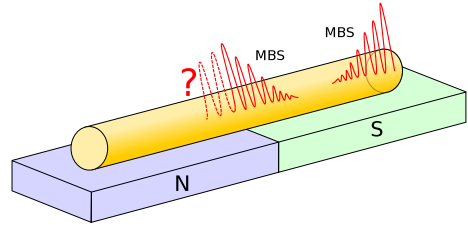}
\caption{
Schematic representation of investigated system. 
A semiconductor nanowire (yellow cylinder) is deposited on two kinds of substrates: normal (N) and superconducting (S). 
Red curve represent our question regarding the effect of Majorana wavefunction \textit{leakage} into the part of the nanowire in normal part of the setup.
}
\label{fig.schem}
\end{figure}

Majorana states emerge on the edges of low dimensional systems. 
However, contrary to the Kitaev toy model~\cite{kitaev.01}, MBS are not localized exactly on the last sites of theoretical nanowire but are spread about the edge as its wavefunction is spread as well~\cite{fleckenstein.dominguez.2017}.
If there exists a part of the nanowire that does not share the topological character with a non--trivial part, the MBS can \textit{leak} into this region even though it does not meet the topologically non--trivial criteria~\cite{deng.vaitiekenas.16,vernek.penteado.14,ptok.kobialka.2017}.
To test this phenomena we propose a nanowire deposited on the surface composed of normal (N) and superconducting (S) regions (Fig.~\ref{fig.schem}).
The part of nanowire deposited on the normal part of substrate acts as an topologically trivial elongation of nanowire due to the absence of superconducting gap $\Delta$.
Such elongation should influence the leakage of MBS wavefunction as it is flowing into the additional region of nanowire, hosting dozens of available states for Majorana wavefunction to leak into~\cite{klinovaja.loss.12,prada.sanjose.12,guigou.sedlmayr.16}.

The main purpose of this paper is to investigate the leakage phenomena of the MBS from one part of the system to another.
We will study this using the system schematically shown in Fig.~\ref{fig.schem} -- a semiconductor nanowire (with strong spin orbit coupling) located partially on the normal/superconducting base.
We study evolution of the MBS by investigation of the local density of states (LDOS) with respect to the electrostatic potential $V_{N}$ on the normal part of system and magnetic field $h$ directed along the wire. 
This paper is constructed as follows: firstly, we described a Hamiltionian model and lay out used techniques in Section~\ref{sec.model}. 
Afterwards, in Section~\ref{sec.num} we present our numerical calculations and interpretations of obtained results.
Finally, we summarize our findings in the Section~\ref{sec.sum}.

\section{Model and technique}
\label{sec.model}

For description of the setup described by Fig.~\ref{fig.schem}, we will use a microscopic model in real space with Hamiltonian  $\mathcal{H} = \mathcal{H}_{wire} +  \mathcal{H}_{SOC} + \mathcal{H}_{prox} + \mathcal{H}_{N}$.
Firstly, we describe the mobile electrons in the wire by:
\begin{eqnarray}
\mathcal{H}_{wire} = \sum_{ij\sigma} \left\lbrace - t \delta_{\langle i,j \rangle} - \left( \mu + \sigma h \right) \delta_{ij} \right\rbrace c_{i\sigma}^{\dagger} c_{j\sigma} ,
\end{eqnarray}
where $c_{i\sigma}^{\dagger}$ ($c_{i\sigma}$) describes creation (annihilation) operator of the electron with spin $\sigma$ in site {\it i}-th, $t$ denotes a hopping integral between the nearest-neighbor sites, whereas $\mu$ is a chemical potential.
Here $h$ describes the magnetic field parallel to the wire in the Zeeman form, which is necessary to the realization of the MBS.
We neglected the orbital effects which have destructive impact on the MBS~\cite{kiczek.ptok.17,nowak.wojcik.18}.
The spin-orbit coupling (SOC) in the whole wire can be expressed by:
\begin{eqnarray}
\mathcal{H}_{SOC} = - i \lambda \sum_{i \sigma\sigma'} c_{i\sigma}^{\dagger} \hat{\sigma}_{y}^{\sigma\sigma'} c_{i+1\sigma'} + h.c. ,
\end{eqnarray}
where $\hat{\sigma}_{y}$ is the second Pauli matrix.

As we mentioned, we are assuming the nanowire deposited on the superconducting (S) and normal (N) substrates (Fig.~\ref{fig.schem}).
In consequence of the proximity effects in a part of nanowire in vicinity of the S, the superconducting energy gap $\Delta$ is induced in the wire in following way:
\begin{eqnarray}
\mathcal{H}_{prox} = \sum_{i \in \text{S}} \Delta \left( c_{i\downarrow} c_{i\uparrow} +   c_{i\uparrow}^{\dagger} c_{i\downarrow}^{\dagger}  \right) .
\end{eqnarray}
Similarly, in the part of the wire in vicinity of the N, the occupation of the nanowire is changed electrostatically by the $V_{N}$ voltage applied to the N part:
\begin{eqnarray}
\mathcal{H}_{N} &=& \sum_{i \in \text{N}} V_{N} \left( c_{i\uparrow}^{\dagger} c_{i\uparrow} + c_{i\downarrow}^{\dagger} c_{i\downarrow} \right) .
\end{eqnarray}

The Hamiltonian $\mathcal{H}$ can be exactly diagonalized by transformation~\cite{ptok.kobialka.2017}
\begin{eqnarray}
\label{eq.bvtransform} c_{i\sigma} = \sum_{n} \left( u_{in\sigma} \gamma_{n} 
- \sigma v_{in\sigma}^{\ast} \gamma_{n}^{\dagger} \right) 
\end{eqnarray}
where $\gamma_{n}$ and  $\gamma_{n}^{\dagger}$ are the quasiparticle fermionic operators, while $u_{in\sigma}$ and $v_{in\sigma}$ is the eigenvectors.
This leads to the Bogoliubov--de~Gennes (BdG) equations~\cite{degennes.89}:
\begin{eqnarray}
\label{eq.bdg} \mathcal{E}_{n} \psi_{in}
= \sum_{j} \left(
\begin{array}{cccc}
H_{ij\uparrow} & D_{ij} & S_{ij}^{\uparrow\downarrow} & 0 \\ 
D_{ij}^{\ast} & -H_{ij\downarrow}^{\ast} & 0 & S_{ij}^{\downarrow\uparrow} \\ 
S_{ij}^{\downarrow\uparrow} & 0 & H_{ij\downarrow} & D_{ij} \\ 
0 & S_{ij}^{\uparrow\downarrow} & D_{ij}^{\ast} & -H_{ij\uparrow}^{\ast}
\end{array} 
\right) 
\psi_{jn} ,
\end{eqnarray}
where $\psi_{in} = ( u_{in\uparrow} , v_{in\downarrow} ,  u_{in\downarrow} , v_{in\uparrow} )^{T}$
Here, the single-particle term $H_{ij\sigma} = - t \delta_{\langle i,j \rangle} 
- \left( \mu + \sigma h \right) \delta_{ij} + \forall_{i \in \text{N}} V_{N} \delta_{ij}$ 
and the spin-orbit coupling term
$S_{ij}^{\sigma\sigma'} = - i \lambda \hat{\sigma}_{y}^{\sigma\sigma'}$.
$D_{ij} = \forall_{i \in \text{S}} \Delta \delta_{ij}$ describes the superconducting gap. 
We employ the eigenvectors and eigenvalues of transformed Hamiltonian $\mathcal{H}$ solving the BdG equations, which allow to calculate the local density of states (LDOS)~\cite{matsui.sato.03}:
\begin{eqnarray}
\nonumber \rho_{i\sigma} ( \omega ) = \sum_{n} \left[ | u_{in\sigma} |^{2} \delta 
( \omega - \mathcal{E}_{n} ) + | v_{in\sigma} |^{2} \delta ( \omega + \mathcal{E}_{n} ) 
\right] . \\
\label{eq.ldos}
\end{eqnarray}


\section{Numerical results and discussion}
\label{sec.num}

In this section we shall describe the physical properties of the investigated system and the phenomena occurring as a result of the voltage $V_{N}$ manipulation.
Calculations has been performed in the system with $300$ sites with fixed
$\mu / t = - 2$, $\lambda / t = 0.15$, $\Delta / t = 0.2$ and $k_{B}T / t = 0$ .

\begin{figure}[!b]
\centering
\includegraphics[width=\linewidth]{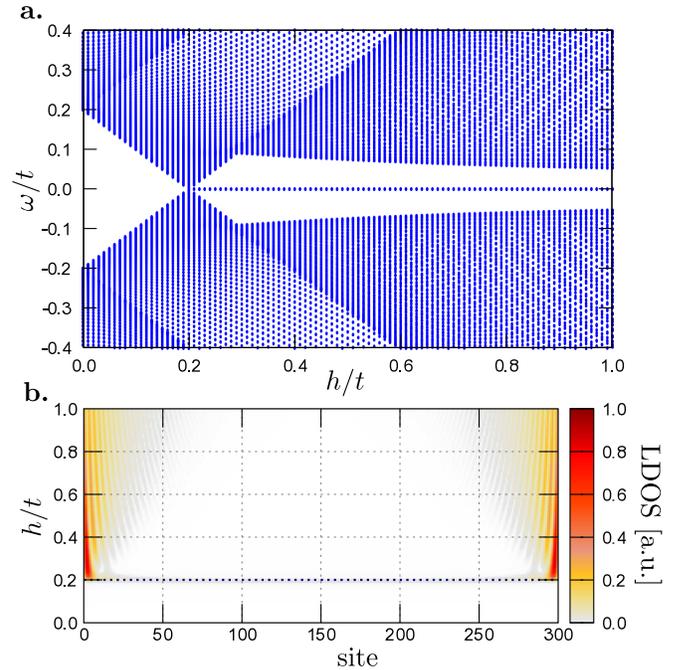}
\caption{
(a) Energy levels of nanowire as a function of magnetic field $h$. 
The zero energy state correspond to the double--degenerate Majorana bound states.
(b) Zero energy LDOS of nanowire on given site, as a function of the magnetic field $h$.
MBS emerge at the ends of the nanowire, as the system undergoes a transition from topologically trivial to non--trivial phase ($h_{c}=0.2t$).}
\label{fig.ldos}
\end{figure}

We shall begin with a brief description of the studied system in the absence of the N part of nanowire.
In the low dimensional system with SOC, superconductivity and magnetic field a topological phase transition can occur as the magnetic field crosses a critical threshold value of $h_{c} = \sqrt{ \Delta^{2} + ( 2 t \pm \mu )^{2} }$~\cite{sato.fujimoto.09,sato.takahashi.09,sato.takahashi.10}.
With the increase of magnetic field value $h$, we obtain the standard Hamiltonian spectrum (Fig.~\ref{fig.ldos}.a). 
For chosen parameters, the critical magnetic field threshold occurs for $h_{c} / t = 0.2$.
For this value of magnetic field ($h = h_{c}$), the gap of the system closes and reopens, therefore changing the topological state of system.
As a result of this, Andreev bound states originating in symmetric (with respect to the Fermi level) energies coalesce at zero energy creating a MBS~\cite{liu.sau.17b,ptok.kobialka.2017,hoffman.schrade.2016}.
Zero energy states correspond to the states localized on the edge of nanowire, which is shown in zero--energy LDOS (Fig.~\ref{fig.ldos}.b).
After a passing through the magnetic field threshold we can observe non-zero LDOS concentrated at the ends of nanowire.
Additionally, increase in Zeeman field \textit{dissolves} this edge state even further, due to the decrease of topological gap and changing the Majorana states oscillation in space.

\begin{figure}[!t]
\centering
\includegraphics[width=\linewidth]{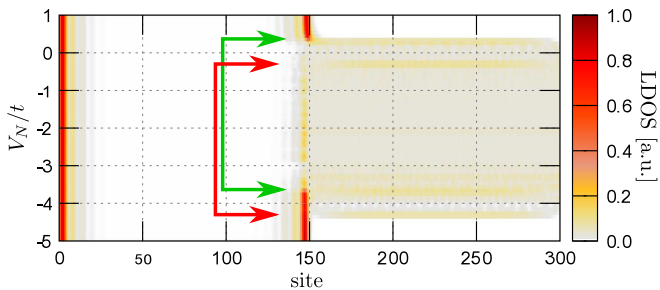}
\caption{
Influence of voltage $V_{N}$ on the normal part of nanowire LDOS at zero energy. 
Energies $V_{N}$ between red (green) arrows correspond to the energies of states with dominant $\downarrow$ ($\uparrow$) spin character.
}
\label{fig.ldos_ns}
\end{figure}

In above calculations we assumed that the superconducting gap $\Delta$ is independent of the magnetic field. 
However, we must have in mind that the experimental venue the superconducting gap changes with magnetic field in a following way $\Delta (h) \simeq \Delta\sqrt{1-(h/h_{c2})^{2}}$~\cite{liu.sau.17b}
As a result, for the magnetic field $h_{c2}$ superconducting gap $\Delta(h)$ closes and the system transitions to the normal state.
Consequently, topological gap vanishes as well and therefore MBS would exist only for $h_{c} < h <  h_{c2}$.
In further results we use constant values of $\Delta$ and $h$, thus condition mentioned above does not influence the calculations.

\begin{figure}[!t]
\centering
\includegraphics[width=\linewidth]{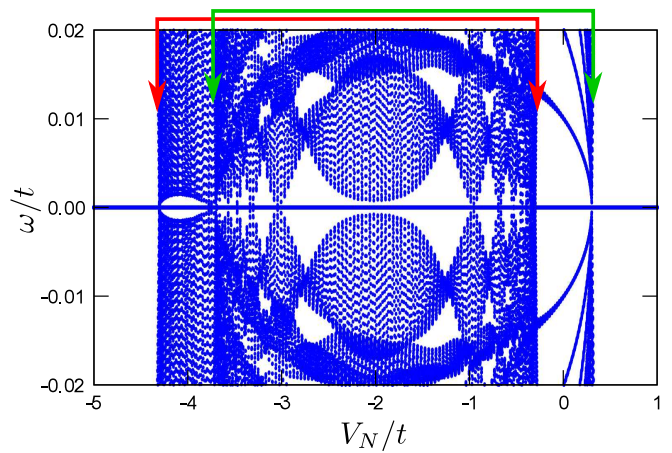}
\caption{
Low energy eigenvalues of the system as a function of the normal part voltage $V_{N}$.
Potential $V_{N}$ between red (green) arrows corresponds to the energies of states with dominant $\downarrow$ ($\uparrow$) spin character.
Outside of that region the only residing states are ABS.
}
\label{fig.warw_ns}
\end{figure}

Now, we will describe the results in a case of the nanowire located partially in the S (sites $\in \langle 1 , 150 \rangle$) and the N (sites $\in \langle 151 , 300 \rangle$) base.
Due to fact that the system has to be under the influence of spin orbit coupling, spin is not a valid quantum number anymore. 
Instead we consider states in terms of spin dominant character.

At the end of the nanowire, for a case of the non--trivial topological phase (Fig.~\ref{fig.ldos}.b), the typical Majorana wavefunction oscillation occurs~\cite{klinovaja.loss.12,hegde.vishveshwara.16}.
Similar results can be observed for any value of voltage $V_{N}$ (Fig.~\ref{fig.ldos_ns}).
This changes drastically near the N/S interface (in the center of the nanowire).
Existence of the interface between normal and superconducting part enables for Majorana wavefunction to \textit{leak} to the rest of nanowire (in our case from left to right). 
Applying a potential to the normal part of the nanowire changes the energy of available states. 
In the range indicated by green arrows potential shifts the energy of normal levels in such manner that some states cross the Fermi level. 
If the majority character of states shifted by voltage coincides with MBS ``spin''~\cite{sticlet.bena.12}, then MBS can \textit{leak} to the normal part of nanowire.
We must have in mind, that the magnetic field $h$ shifts the energy of the $\uparrow$/$\downarrow$ states in the N part. 
In consequence, the states with $\uparrow$ ($\downarrow$) dominant character are located between green (red) arrows on $V_{N}$ in Fig.~\ref{fig.ldos_ns}.
Moreover, in regions below green (above red) arrows of the potential $V_{N}$, only the states with $\downarrow$ ($\uparrow$) spin dominant character are available at the zero energy. 
At the same time, dominant spin in the S region is still $\uparrow$ (for $h > 0$).
In consequence of this, for $V_{N}$ around $0 t$ (between green and red upper arrows) we can observe {\it leakage} of the MBS from the left (S) region to the right (N) region.
This behavior are not observed for $V_{N}$ around $- 4 t$.

Settings described above correspond to the eigenvalues presented on Fig.~\ref{fig.warw_ns}.
We can see the interplay between zero-energy MBS and in-gap ABS states originating in nanowire.
Significant asymmetry of ABS as a function of potential $V_{N}$ can be observed  in regions between red and green arrows. 
This is a result of the availability of spin dominant levels aligned (left region) and misaligned (right region) with MBS spin and therefore can lead to MBS \textit{leakage} (in the left region between red and green arrows).

\begin{figure}[!t]
\centering
\includegraphics[width=\linewidth]{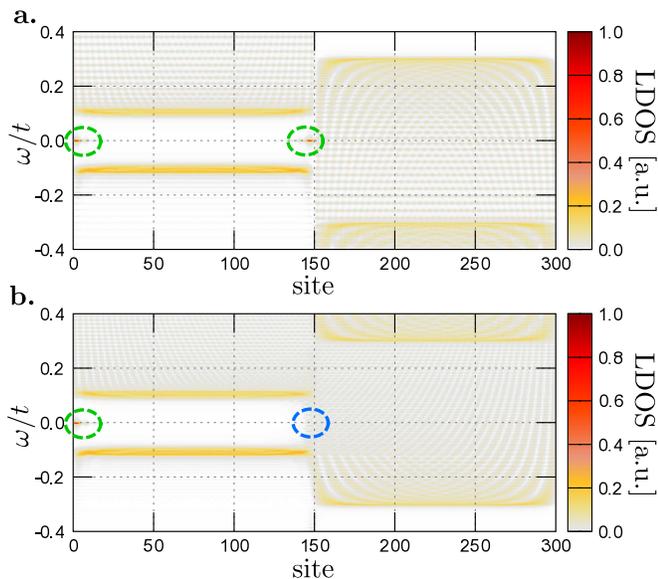}
\caption{
LDOS of nanowire for $V_{N}$ equal $-4 t$ (a) and $0 t$ (b).
}
\label{fig.ldos_vg}
\end{figure}

The MBS emerge in the system for any  value $V_{N}$.
However, its localization at the end of the S part is changed. 
This can be seen explicitly in a LDOS along the whole wire for chosen $V_{N}$ (Fig.~\ref{fig.ldos_vg}).
In a case of the spin $\downarrow$ majority character (panel a), MBS is stationary localized at the ends of the topologically non--trivial part of the nanowire (green ellipses).
MBS \textit{leaks} only very slightly into the states of opposite spin majority, therefore the LDOS around the end of S part of nanowire does not diminish drastically.
We observe two symmetrical MBS with comparable spectral weight in the S region.
Contrary to that (panel b), only one of the MBS can be clearly distinguished (green ellipse), while second one is delocalized along N part of the wire (blue ellipse).
This is a result of \textit{leakage} of one of the MBS to the spin $\uparrow$ majority normal nanowire state and can be understood as the enhancement of the LDOS at $\omega=0t$ with qualitatively denser distribution of states than on the Fig.~\ref{fig.ldos_vg}.a.
The normal part of system plays a role of an ``new'' edge of topologically non--trivial part and behaves like such enabling the Majorana wavefunction oscillation in its region.
As a consequence, spectral weight of right MBS is partially transferred to the normal part of wire, creating the pairs of the MBS with strongly non--symmetric spectral weight (cf. green and blue ellipses at panel Fig.~\ref{fig.ldos_vg}.b).

\section{Summary}
\label{sec.sum}

Recent experimental studies introduced the phenomenon of MBS \textit{leakage}, due to the spreading of the Majorana wavefunction along the wire to the attached structures ~\cite{deng.vaitiekenas.16}.
It shows an interesting opportunity for both fundamental studies and application in quantum computing due to the ease of manipulation of the potential on quantum dot~\cite{ptok.kobialka.2017,prada.aguado.17}.
In this paper we ask a question regarding conditions of the leakage and how does it correspond to the localization of MBS on the ends of topologically non--trivial nanowire.
We proposed a system to test this feature.

In typical situation, the MBS exhibits spatial oscillations in LDOS spectrum related to the oscillations of Majorana wavefunction in space. 
Tuning the chemical potential of normal part of the nanowire by electrostatic means, shifts the states on the Fermi level of the spin dominant character aligned with the MBS ``spin'' and can lead to the \textit{leakage} of one of the bound states from non-trivial to topologically trivial  part of the wire. 
Moreover, the localization of the zero-energy state is diminished sufficiently enough to be indistinguishable from the background zero energy ABS (in trivial part of nanowire).
This creates a system of effective topological nanowire with strongly non--symmetric spectral weight, where only one of the MBS can be clearly distinguished due to the severe leakage of the other MBS to the normal part of the system.
We strongly believe that the experimental realization of this type system can be helpfully in distinguish trivial (Andreev) and non-trivial (Majorana) zero energy bound states.

\begin{acknowledgments}
This work was supported by the National Science Centre (NCN, Poland) under grants DEC-2014/13/B/ST3/04451 (A.K.) and UMO-2017/25/B/ST3/02586 (A.P.).
\end{acknowledgments}

\bibliography{biblio}

\end{document}